\documentclass[11pt,english]{article}
\usepackage[T1]{fontenc}
\usepackage[latin9]{inputenc}

\makeatletter
\usepackage{hyperref}\pdfoutput=1
 \usepackage{hyperref}
 \pdfoutput=1

\makeatother

\usepackage{babel}

\begin{document}

\title{Stretching Folding Instability and Nanoemulsions}

\author{Chon U Chan, Claus Dieter Ohl\\
 \\
\vspace{6pt}
 School of Physical and Mathematical Sciences (SPMS)\\
 Nanyang Technological University\\
 Division of Physics and Applied Physics\\
 Singapore 637371}

\maketitle

\begin{abstract}
Here we show a folding-stretching instability in a microfluidic flow
focusing device using silicon oil (100cSt) and water. The fluid dynamics
video demonstrates an oscillating thread of oil focused by two co-flowing
streams of water. We show several high-speed sequences of these oscillations
with 30,000 frames/s. Once the thread is decelerated in a slower moving
pool downstream an instability sets in and water-in-oil droplets are
formed. We reveal the details of the pinch-off with 500,000 frames/s.
The pinch-off is so repeatable that complex droplet patterns emerge.
Some of droplets are below the resolution limit, thus smaller than
1 micrometer in diameter.
\end{abstract}

Two files are available a
\href{http://ecommons.library.cornell.edu/bitstream/1813/14083/4/APSDFD2009_Chan_Ohl_high_resolution.mpg}{low resolution}
and a 
\href{http://ecommons.library.cornell.edu/bitstream/1813/14083/3/APSDFD2009_Chan_Ohl_low_resolution.mpg}{high resolution} video.

\end{document}